\documentclass{iau}
\usepackage{graphicx}

\title[Decolonizing technosignature] 
{Decolonizing technosignatures: Reimagining technosignatures through
an Indigenist lens}

\author[Hilding R.~Neilson]   
{Hilding R.~Neilson$^1$
 }

\affiliation{$^1$Department of Physics \& Physical Oceanography, \\ Memorial University of Newfoundland \& Labrador \\ St.~John's, Newfoundland \& Labrador, Canada \\ email: {\tt hneilson@mun.ca} }

\pubyear{2026}
\volume{404}  
\pagerange{119--126}
\jname{IAUS 404: Advancing the Search for Technosignatures
}
\editors{???, eds.}
\begin{document}

\maketitle

\begin{abstract}
There exist numerous proposals for potential technosignatures that we can search for, but they all tend to be deviations from possible naturally occurring signals.  That is, technosignatures tend to be a search for pollution. Following a number of works by Indigenous scholars, pollution is colonization, even if that pollution is light-years away.  In this situations we are searching for colonial societies that are a reflection of colonial societies on Earth. This may help explain the Fermi Paradox.  We are searching for systems that are highly polluting and potentially destructive, while we do not search for extraterrestrial societies that are less impactful on their celestial environment.  This also suggests that we might consider technosignatures that act to support natural environments and might be more closely related to biosignatures, that is a technology that reflects a supportive relationship with nature.  

\end{abstract}

\firstsection 
\section{Introduction}
The hunt to detect technosignatures, or signs of so-called advanced civilizations, has evolved significantly since Dr. Frank Drake began using radio telescopes to search for radio transmissions from other stars (\cite{Drake1962}).  Since then, there have been many ideas for potential technosignatures from megastructures to civilization-produced industrial products that would be evidence of advanced civiliizations.

However, technosignatures have to be confirmed to be independent of biological and natural chemical signatures so that there is little possibility that these technosignatures are real and not being confused by other sources.  In this situation, technosignatures must exist at both global scales and must be significantly different from natural processes for us to potentially be able to detect them with our current facilities.  In short, technosignatures are mostly a measure of pollution in other worlds, be it industrial products or light pollution or radio broadcastings, etc.  Even megastructures such as Dyson spheres (\cite{Dyson1960}) can be considered polluting as they both require destruction of solar systems and they block radiation  from the host star that would interact with the interstellar environment.  In short, the current dominant view of technosignatures is one of pollution.

If we consider that technosignatures are pollution, then we should also consider how our understanding of technosignatures is born through Western Sciences and methodologies.  As such, I argue in this work that the current view of technosignatures is limited to a Western Science culture and is, in fact, colonial in nature.  In the next section, I briefly discuss Indigenous methods of science followed by how we, as astronomers, tend to have a colonial model of technosignatures and I conclude with a discussion of how we might begin to decolonize technosignatures.

\section{Indigenous methods and sciences}
When we consider Indigenous knowledges, particularly of peoples in what today  is called  Canada and the United States, I start by acknowledging that Indigenous peoples have lived in their territories since time immemorial.  I also acknowledge that I am of Mi'kmaw First Nation and Settler identity. This means that Indigenous peoples have lived under the same night sky and on the same land for millennia, observing, learning and building an understanding of the natural phenomena all around.  This includes understanding the land and sky from planets to stars to transients.  This relationship with the Land (capitalized here to reflect its animate nature).  Some of the key points, as described by  \cite{ Battiste2013,Cajete2000,  Lipe2019, Smith2021} and others,  are:
\begin{itemize}
\item What is above reflects below. 
\item Knowledge is relational.
\item Knowledge is (w)holistic   
\item Nature is sacred and familial and is kin. 
\end{itemize}
This is not a comprehensive list of Indigenous ways of knowing, but are key aspects for consideration in this work.  It must also be noted that there are no one Indigenous knowledges or ways of knowing, different communities and nations have their own ways of knowing. These Indigenous ways of knowing are given as a general set on commonalities.  

These Indigenous ways of knowing differ from traditional Western methods of science in that Western methods focus on the concepts (see for example, \cite{Shapin2018}):
\begin{itemize}
\item Knowledge is objective
\item Knowledge is in disciplines
\item Nature is hierarchic.
\end{itemize}
These perspectives clearly differ.  For instance, the idea that knowledge is objective is a part of the scientific method and means that under the same conditions people conducting the same observations or experiments should record the same results.  For Indigenous methods, the concept that knowledge is relational means 1) that we have a responsibility for the knowledge we learn, and 2) that any knowledge also depends on the person or persons undertaking the experiment.  Another way to phrase this is that the idea that knowledge is objective means there is an objective truth while knowledge is relational means there can be many truths. 

The second concept highlights how knowledge is Western Science tends to be structured in disciplines. Universities tend to be constructed in separate departments and faculties, such as physics, math or social sciences. The knowledges created in these fields tend to be held in isolation and when there is cooperation between researchers in disparate fields we celebrate their interdisciplinarity (\cite{Shapin2018}).  On the other hand Indigenous knowledges are (w)holistic in that knowledges and stories include connections between animals, Nature, humans, and the night sky.  We see in this Indigenous stories across many nations (e.g. \cite{Bartlett, Buck, MacDonald}). 

The third concept refers to how the natural world arounds us tends to be viewed. In Western Science, nature is a hierarchy in that humans consider themselves the apex of nature, above animals, plants, water, and air (\cite{Jensen2016, Kimmerer2013, Noon2022}).  In this way, science is driven by an anthropocentric view that is centred on Western knowledges.  Conversely, for many Indigenous knowledges, Nature is sacred and is familial.  In this way, animals, plants, water and air can be considered kin and have inherent rights of their own. 

When we consider these concepts together then we can see the Indigenous Ways of Knowing can be very different than traditional Western Science methods and as such offer perspectives on nature and phenomena that can be missed when we consider only Western Science.  One way to think about this is until less than a century ago the field of astronomy was only understood at optical wavelengths.  When telescope at radio or ultraviolet wavelengths were developed, astronomers began to see a very different Universe and bringing observations from many different wavelengths together we could build a greater view of the Universe. 

\section{Are technosignatures pollution?}
When we consider technosignatures through Indigenous methods a number of issues become apparent. The first is that the genesis of many proposed technosignatures come from observations of Western Societies and the interactions of technologies with the environment.  This could be from radio emissions being broadcast into outer space or optical light pollution being wasted.  It might include particulate matter like CFCs, or be related to smog.  The second issue is that a number of proposed technosignatures are based on science fiction that is also born from Western Science perspectives.  Notably this would include Dyson Spheres, interstellar probes, lasers and other technologies that could born by alien civilizations that operate at interstellar scales.  However, these concepts, while not exhaustive of all technosignature literature, tend to derive from the concepts of Western Science.

As such, technosignatures also have the property that the signal is a deviation from natural signals. If there is no deviation then we cannot detect an alien civilization.  Furthermore, with the currently technology on Earth, we would require that a technosignature have a significant deviation from natural signals for be confirmed.  As such these technosignatures tend to be forms of pollution in that they impact the natural environment in some way either by extraction in their creation and maintenance or via impacts on the environment around them.  These impacts are part of a Western Science methodology where the creators of the technology consider themselves the apex of their world and environment. 

This differs from what might be expected if we consider Indigenous methods.  Since Indigenous methodologies consider all beings as having inherit rights and as equals be it animals, water, air, etc. then this means any created technologies must be supportive of nature and carry our relationships. Therefore, technologies that are consistent with this responsibility would not offer a significant harm or negative impacts on the environment, hence would be unlikely to create a significant and detectable technosignature like those consistent with Western Science.  

In this situation, technological civilizations that we can detect would be consistent with Western Societies and as such would be born from colonial perspectives.  \cite{Liboiron2021} noted that pollution is colonialism, particularly if it is wasteful and damaging to non-human relations without consideration. In this way, we should consider that technosignatures, in their current formulation, are colonialism and, as such, astronomers are only looking for a specific subset of technosignatures that represent only a specific type of alien civilization.  

This result can be considered through the lens of the Fermi Paradox.  If we are only looking for the most polluting technosignatures then one might expect these are the most likely to be self-destructive or perhaps this is a short-lived transition to a more sustainable interaction with the environment (planetary and space) that would be more challenging to detect. As such, we might expect that the technosignatures we search for represent unlikely possibilities, hence is connected to various proposed resolutions to the Fermi Paradox (\cite{Webb2015}).

\section{Conclusions}
In this work, I argue that the current perspective of searching for extraterrestrial technologies is colonial in practice given the focus on searching for pollution and deviations from natural signatures.  As such, it is important that as a research field, we seek to decolonize technosignatures. Given the current ability to detect and understand technosignatures, it is not obvious how the research field might accomplish this.  Given that technologies should  support and live in relation with Nature when considered through Indigenous methods, then technologies will not likely offer signatures that are detectable using the current technologies and methods in research astronomy.  

However, having noted this issue we might consider what might signatures look like from these technologies and civilizations.  One consideration is what if technology supports and enhances natural signatures.  Could we have a technosignature that mimics biosignatures but is stronger than what might be expected for a given system? A second consideration is what if technosignatures that might be a deviation from natural signatures are less detectable because it is recognized that they cannot be developed at a scale that we might detect.  In this situation, these technosignatures would be far more challenging for current observers but also might be more common as they would also be less destructive.  In the end, we, as a research community, should expand our view of technosignatures being the traditional Western and colonial lens and be more inclusive of Indigenous methods.  By learning from Indigeneous methods we can apply the philosophy of Two-Eyed Seeing (\cite{Bartlett2012}) that uses Western Science as one lens and indigenous methods as a second lens to build a deeper and more profound view of technosignatures and the evolution of life in the Galaxy.

\end{document}